\documentclass[a4paper, amsfonts, amssymb, amsmath, reprint, showkeys, nofootinbib, twoside]{revtex4-1}
\usepackage[english]{babel}
\usepackage[utf8]{inputenc}
\usepackage[colorinlistoftodos, color=green!40, prependcaption]{todonotes}
\usepackage{amsthm}
\usepackage{mathtools}
\usepackage{physics}
\usepackage{xcolor}
\usepackage{graphicx}
\usepackage[left=23mm,right=13mm,top=35mm,columnsep=15pt]{geometry} 
\usepackage{adjustbox}
\usepackage{placeins}
\usepackage[T1]{fontenc}
\usepackage{lipsum}
\usepackage{csquotes}

\usepackage[pdftex, pdftitle={A New World Framework}, pdfauthor={Dacheng Zhou}]{hyperref} 
\begin{document}
\title{A New World Framework: The Three Realms and Six Layers Model}

\author{Dacheng Zhou}
    \email[Correspondence email address: ]{zhou_d_c@hotmail.com}
    \affiliation{University of Sci. and Tech. of China, Institute of Physics, Hefei, Anhui, China}

\date{\today} 

\begin{abstract}
This paper introduces an innovative framework for understanding the world, termed the "Three Realms and Six Layers Model". Based on the concept of scale, the world is divided into three realms, each encompassing six layers, with a ten-thousand-fold difference in scale between adjacent layers. This unique division reveals the variations in laws at different levels and the fundamental changes in laws when transitioning between realms. The model offers a new perspective for addressing interdisciplinary issues, especially in understanding the behavior of large-scale systems and their connection to microscopic phenomena. In summary, the "Three Realms and Six Layers Model" provides a novel tool for comprehending the diversity and complexity of the universe.
\end{abstract}


\maketitle

\section{Introduction} \label{sec:introduction}
In exploring this intricate world, we continuously seek connections and distinctions between various disciplines, trying to unveil the secrets of nature through a deep understanding of fundamental subjects. However, as research progresses, we find that the perspective of a single discipline often struggles to fully explain the complexity of phenomena. This raises a fundamental question: Is there a method that can transcend different disciplines, allowing for a simple yet effective classification and understanding of them?

In 1972, Nobel Laureate in Physics, Philip W. Anderson, in his groundbreaking paper "More is Different"\cite{anderson1972more},
 proposed a thought-provoking view: Different physical laws exist at different scales in nature. Anderson emphasized that higher-level phenomena cannot be fully explained by more basic physical laws alone, revealing the limitations of reductionist approaches. His viewpoint not only challenged traditional thinking in the physics community but also provided theoretical support for interdisciplinary research.

Inspired by Anderson, this paper proposes a new world model, the "Three Realms, Six Levels Model." This model aims to define and explain the different laws of objects at various scales, from the microscopic to the macroscopic, from basic physics to complex life phenomena, attempting to construct a framework to unify and simplify our understanding of the world. By dividing the world into different realms and levels, we can see more clearly how new laws and phenomena emerge with changes in scale, and how these laws and phenomena interact and transform. This not only helps us better understand the diversity and complexity of the natural world but also offers a new perspective and tool for interdisciplinary research.

This paper will first review Anderson's theory, then detail the construction process and main content of the "Three Realms, Six Levels Model", and finally explore the model's application prospects and potential value across different disciplines. Through this research, we hope to provide a new methodological framework for understanding and classifying the complex world.

Therefore, it can be hypothesized that with each certain degree of scale increase, a new symmetry breaking occurs, leading to the emergence of new laws. This paper proposes a scale-based approach to categorize the different levels of the universe, aiming to measure the symmetry breaking at various scales.
 
\section{Assumptions} \label{sec:assumptions}
Assuming the following conditions:

1. Define \textbf{a layer} as a specific range of scales, for example, from 1mm to 10m. If an object's size falls within this range, then the object belongs to this layer;

2. The maximum scale of each layer is 10,000 times its minimum scale, for example: 10 meters = 10,000 * 1 millimeter;
   
3. The layers are continuous, meaning the maximum scale of one layer equals the minimum scale of the next layer. Thus, as long as there are enough layers, any object can be classified into a certain layer;
   
4. Considering that the Planck length (1.6E-35 meters) is theoretically the smallest unit of length, it is therefore set as the minimum scale at the most fundamental level. For the sake of simplifying calculations, it can also be simplified to 1E-35 meters, which currently does not seem to affect the outcome;
   
5. Define \textbf{a realm} contains six continuous layers. Therefore, within a realm, the difference between the maximum and minimum lengths is \(10,000^6 = 10^{24}\).

The scale of each layer, \(L\), can be represented by the following formula:
$$L_{\text{start}} = 10^{24r+4l-15}$$
$$L_{\text{end}} = 10^{24r+4l-11}$$
$$l \in \{1,2,3,4,5,6\} , r \in \{-1,0,1\}$$
Where \textbf{l} is the layer number, and \textbf{r} is the realm number.

\bigskip
Based on the above assumptions, we arrive at the following conclusions:

1. The smallest realm has a scale range from 1E-35 to 1E-11 meters. Its upper limit of 1E-11 meters is close to the size of a hydrogen atom, namely, the Bohr radius, about 5E-11 meters. Since the main laws in this scale range are quantum mechanics, it is referred to as the Quantum Realm, or mythologically as the "Underworld";

2. Above the Quantum Realm, the range from 1E-11 meters to 1E13 meters covers the world as we know it, with the smallest structure being the atom. Therefore, it is referred to as the Atomic Realm, or "Human Realm";
   
3. The size of the observable universe is approximately 1E27 meters, and thus the range from 1E13 meters to 1E37 meters (far beyond the size of the observable universe) constitutes another realm. Its smallest unit is the galaxy, hence it is referred to as the Galactic Realm, or "Star Realm".

Through this classification method, we can divide the entire universe into the Underworld (Quantum Realm), Human Realm (Atomic Realm), and Star Realm (Galactic Realm).

\section{The Laws of Each Realm} \label{sec:realms}

In nature, four fundamental forces have been identified: the strong nuclear force, the weak nuclear force, the electromagnetic force, and gravity. These forces exhibit different characteristics on different scales:

1. Strong and weak nuclear forces have a range of action limited to below 1E-15 meters, meaning they only operate within the Underworld (Quantum Realm);

2. In theory, the electromagnetic force has no limit to its range of action. However, due to the neutralization effect between charges, large-scale objects (such as planets) usually appear electrically neutral. Therefore, at the scale of planets or larger, electromagnetic forces are generally not considered, meaning electromagnetic interactions mainly occur within the Human Realm (Atomic Realm) and smaller scales;

3. Gravity also has no limit to its range of action, but when dealing with the interactions between microscopic particles, its influence is usually neglected due to the extremely small mass. Therefore, gravity is primarily considered within the Star Realm (Galactic Realm) and the Human Realm.

As the scale changes, each realm exhibits a dominant force, and there are fundamental equations specifically describing the behavior of objects:

1. The Underworld is dominated by the Schrödinger equation and quantum mechanics;
$$i\hbar\frac{\partial}{\partial t}\Psi(\mathbf{r},t) = \hat{H}\Psi(\mathbf{r},t)$$

2. The Human Realm is dominated by Maxwell's equations and electromagnetism;
$$\nabla \cdot \mathbf{E} = \frac{\rho}{\varepsilon_0}, \quad
\nabla \cdot \mathbf{B} = 0$$
$$\quad
\nabla \times \mathbf{E} = -\frac{\partial \mathbf{B}}{\partial t}, \quad
\nabla \times \mathbf{B} = \mu_0\mathbf{J} + \mu_0\varepsilon_0\frac{\partial \mathbf{E}}{\partial t}$$

3. The Star Realm is dominated by Einstein's field equations and general relativity.
$$G_{\mu\nu} + \Lambda g_{\mu\nu} = \frac{8\pi G}{c^4} T_{\mu\nu}$$

This discovery reveals the applicability and limitations of different theories at different scales:

1. Quantum mechanics is no longer the dominant force in the Human Realm and Star Realm;

2. Electromagnetism no longer plays a dominant role in the Star Realm, and in the Underworld, it combines with quantum mechanics to form Quantum Electrodynamics (QED);

3. Relativity degenerates into classical Newtonian mechanics in the Human Realm and is almost negligible in the Underworld.

Therefore, we can see that these fundamental equations are either ignored in other realms due to their minimal impact or transformed into new forms due to the influence of forces from other realms.

\section{The Laws of Each Layer} \label{sec:layers}

To differentiate between the various layers in our model, we have adopted a naming method that combines the name of the realm with the layer's number (for example: Human-3, Underworld-6). The first layer is at the bottom, ascending to the sixth layer.

Below, we describe the layers understood by humanity.

\bigskip
In the Human Realm (Atomic Realm):

1. Human-1 (0.01 nanometers to 0.1 micrometers): This layer encompasses scales from atoms, molecules to polymers, serving as the stage for chemical reactions and molecular aggregation. Hence, \textbf{chemistry} is the primary discipline describing the laws of this layer;

2. Human-2 (0.1 micrometers to 1 millimeter): This scale includes cells and the vast majority of microorganisms, making it a focus of microbiology and \textbf{biology} research;

3. Human-3 (1 millimeter to 10 meters): Covering most macroscopic life, including humans, the dominant disciplines here are various \textbf{technologies}, focusing on how humans use tools to transform the world;

4. Human-4 (10 meters to 100 kilometers): This range includes the scale of human societies and animal groups, with \textbf{sociology} and related humanities (such as history, philosophy, arts, politics, etc.) being the main disciplines of this layer;

5. Human-5 (100 kilometers to one million kilometers): Involving geological structures and formations, \textbf{geography} becomes the primary discipline for studying this layer;

6. Human-6 (100 million meters to one trillion meters): This is the scale of galactic structures, with \textbf{astronomy} being the dominant discipline.

 \bigskip
In the Underworld Realm (Quantum Realm):

1. Underworld-6 (1E-15 to 1E-11 meters): Includes the scale of protons, neutrons, and other atomic nuclei, with \textbf{nuclear physics} (quantum mechanics) being the main discipline;

2. Underworld-5 (1E-19 to 1E-15 meters): Encompasses the scale of fundamental particles like electrons and quarks, with \textbf{particle physics} (including the Standard Model, QED, QCD, etc.) being the leading discipline;

3. Underworld-1 to 4: These layers exceed the current exploratory capabilities of humanity.

 \bigskip
In the Star Realm:

1. Star-1 (1E13 to 1E17 meters): Includes galactic structures such as the Kuiper Belt and the entire Solar System, with astronomy continuing to play a significant role;

2. Star-2 (1E17 to 1E21 meters): Covers structures like galaxy filaments that comprise multiple galactic systems;
   
3. Star-3 (1E21 to 1E25 meters): Includes galaxies, galaxy clusters, and superclusters;

4. Star-4 (1E25 to 1E29 meters): Covers superstructures and the scale of the observable universe;

5. Star-5 and 6: Similarly, exceed the current exploratory range of humanity.

Through this hierarchical division, we can clearly define the characteristics and dominant disciplines of each scale level, providing a clear framework for research in different fields.

Each level offers a unique perspective for human understanding of the world. Perhaps we can regard the patterns provided by each level as a kind of Tao. In this way, there are six types of Tao within each realm.

For each level's Tao, patterns may include: 1. Specializing in one Tao boosts efficiency; 2. Adjacent Taos' effectiveness in neighboring levels may lead groups to form dominant Taos spaced two levels apart and suppress others.
\section{Conclusions} \label{sec:conclusions}

In the preceding discussion, we explored the different layers of the universe and their corresponding dominant disciplines. This discovery reveals that within each layer available for study, there exist distinct and systematic laws. Specifically, from particle physics and nuclear physics in the Underworld to chemistry, biology, technology, and sociology in the Human Realm and finally to astronomy in the Star Realm, each layer has its unique field of research and discipline. This layered approach not only demonstrates the orderliness of the natural world but also provides us with a method to simplify the complexity of the world.

Through this layered method, we can understand the structure and operational mechanisms of the universe more systematically. The dominant discipline of each layer offers the best explanation for the laws within that level. This classification not only helps us better organize knowledge, but also provides a framework for interdisciplinary research.

Furthermore, this layered law might also offer guidance for future research. Take dark matter and dark energy as examples; these two concepts remain unsolved mysteries in physics, yet their existence is inferred based on observational data at the scale of the Star Realm. This leads to a hypothesis: dark matter and dark energy might be laws that only manifest in Star-2 and Star-3, and our incomplete understanding of these layers might make these phenomena appear as undiscovered laws. If this hypothesis holds, then a deeper investigation into the different layers of the Star Realm could be key to unlocking the mysteries of dark matter and dark energy.

The model also has its limitations. In this model framework, using 10,000 as the scale multiplier carries a certain subjectivity and is not derived from precise calculations but rather an estimated value. At the same time, it has not been clearly explained why a new set of rules appears when the scale changes by approximately 10,000 times, nor why a completely new set of rules emerges after crossing six layers. Furthermore, whether there are other layers beyond these three known realms remains an unanswered question.

These are puzzles for which answers have not yet been found.

\bibliographystyle{plain}
\bibliography{references}

\end{document}